\documentclass[aps,prd,twocolumn,superscriptaddress,floatfix,notitlepage]{revtex4-1}

\usepackage{amsmath}
\usepackage{amssymb}
\usepackage{dcolumn}
\usepackage{epstopdf}
\usepackage{extarrows}
\usepackage{graphicx}
\usepackage{lineno}
\usepackage{listings}
\usepackage{makecell}
\usepackage{multirow}
\usepackage{xcolor}
\usepackage{yhmath}
\usepackage[colorlinks,citecolor=blue,linkcolor=blue,urlcolor=blue,anchorcolor=blue]{hyperref}
\allowdisplaybreaks
\newcommand\dx{{\rm d}}

\renewcommand\apj{Astrophys. J.}

\newcommand\cqg{Classical Quantum Gravity}

\newcommand\jcap{J. Cosmol. Astropart. Phys.}

\begin{document}

\title{Non-full equivalence of the four-dimensional Einstein-Gauss-Bonnet gravity and Horndeksi gravity for Bianchi type I metric}
\author{S. X. Tian}
\email[]{tshuxun@whu.edu.cn}
\affiliation{Department of Astronomy, Beijing Normal University, 100875 Beijing, China}
\author{Zong-Hong Zhu}
\email[]{zhuzh@whu.edu.cn}
\affiliation{Department of Astronomy, Beijing Normal University, 100875 Beijing, China}
\affiliation{School of Physics and Technology, Wuhan University, 430072 Wuhan, China}
\date{\today}
\begin{abstract}
  The four-dimensional Einstein-Gauss-Bonnet (4DEGB) gravity is proposed as a singular limit of the higher-dimensional EGB gravity with a rescaled coupling constant. Follow-up work indicates that the 4DEGB gravity may be equivalent to a special Horndeski gravity. However, the full equivalence of the two theories has not been proven. In this paper, we test the possible equivalence for Bianchi type I metric. We consider two symmetry choices of the extra dimensions in the 4DEGB gravity and obtain the explicit cosmological evolution equations of the theories. Our result shows that, for one symmetry choice, there is a correspondence between the 4DEGB gravity and the special Horndeski gravity. However, for the other symmetry choice, there is no correspondence between the two theories. Therefore, the 4DEGB gravity is not equivalent to the special Horndeski gravity for the general case.
\end{abstract}
\maketitle

\section{Introduction}\label{sec:01}
Recently Glavan and Lin \cite{Glavan2020.PRL.124.081301} proposed a dimensional-regularization approach to extract the contribution of the Gauss-Bonnet term in four-dimensional spacetime, and bypass Lovelock's theorem in their minds. They started from the action $S=\int\dx^4x\sqrt{-g}(R-2\Lambda+\alpha\mathcal{G})/2\kappa+S_{\rm m}$, where $\mathcal{G}=R^2-4R_{\mu\nu}R^{\mu\nu}+R_{\mu\nu\rho\sigma}R^{\mu\nu\rho\sigma}$ is the Gauss-Bonnet invariant, and the field equation is
\begin{align}\label{eq:01}
  \kappa T^\mu_{\ \nu}&=G^\mu_{\ \nu}+\Lambda\delta^\mu_{\ \nu}
  +\alpha\left(2RR^\mu_{\ \nu}-4R^\mu_{\ \alpha}R_\nu^{\ \alpha}\right.\nonumber\\
  &\quad\left.-4R^\mu_{\ \alpha\nu\beta}R^{\alpha\beta}+2R^\mu_{\ \alpha\beta\gamma}R^{\ \alpha\beta\gamma}_{\nu}
  -\delta^\mu_{\ \nu}\mathcal{G}/2\right).
\end{align}
Lovelock's theorem means that the $\alpha$-term equals to zero in four dimensions. To extract nonzero contribution of the $\alpha$-term, the authors in \cite{Glavan2020.PRL.124.081301} rescaled the coupling constant $\alpha=\tilde{\alpha}/(D-4)$ and defined the new theory as the limit $D\rightarrow4$ of the higher-dimensional gravity. They tested this idea with the flat FLRW cosmology and the static spherically symmetric black hole. A common feature of these two cases is spherical symmetry.

One key ingredient of the above dimensional-regularization approach is to assume the symmetry of the extra dimensions. As pointed out in \cite{Ai2020.CTP.72.095402}, this symmetry is automatically assumed for spherically symmetric cases discussed in \cite{Glavan2020.PRL.124.081301}. However, the symmetry assumption needs to be explicitly given for non-spherically symmetric cases. Generally speaking, different symmetry choices give different evolution equations (see specific examples in Sec. \ref{sec:03}). \cite{Glavan2020.PRL.124.081301} declared that the extra dimensions have no physical meaning, and are only used to define the singular limit. However, the system evolution depends on the choice of symmetry. In this sense, the extra dimensions have physical meaning. Therefore, the theory proposed in \cite{Glavan2020.PRL.124.081301} is essentially a higher-dimensional theory, and does not bypass Lovelock's theorem. To be consistent with the conventions in the literature, we still refer to the theory proposed in \cite{Glavan2020.PRL.124.081301} as the four-dimensional Einstein-Gauss-Bonnet (4DEGB) gravity hereafter.

As we discussed before, the dimensional-regularization approach performed in \cite{Glavan2020.PRL.124.081301} rely on the embedding of the extra dimensions. Introducing an auxiliary scalar field, there is another way to regularize the divergent rescaled action \cite{Mann1993.CQG.10.1405}. The resulting theory belongs to the Horndeski class of scalar-tensor theories, and does not rely on the embedding of the extra dimensions \cite{Fernandes2020.PRD.102.024025,Hennigar2020.2004.09472,Easson2020.2005.12292}. Similar result can also be obtained by Kaluza-Klein reduction with a flat internal space \cite{Lu2020.2003.11552,Kobayashi2020.JCAP.07.013}. As discussed in \cite{Fernandes2020.PRD.102.024025}, all the solutions of the original 4DEGB gravity found in \cite{Glavan2020.PRL.124.081301} are also solutions of the corresponding Horndeski gravity. However, the full equivalence of the two theories has not been proven. In this paper, based on Bianchi type I metric, which describes a homogeneous but anisotropic universe, we provide a counterexample for this possible equivalence. More precisely, we find a solution of the original 4DEGB gravity with one specific symmetry of the extra dimensions does not match the result obtained in the Horndeski gravity.

\section{Horndeski gravity}\label{sec:02}
In this section, we present the cosmological evolution equations of the homogeneous but anisotropic universe described by Bianchi type I metric for the Horndeski gravity. The $D$-dimensional Bianchi type I metric reads $\dx s^2=-c^2\dx t^2+\sum_{i=1}^{D-1}a_i^2\dx x_i^2$, where $a_i=a_i(t)$. The energy-momentum tensor is $T^\mu_{\ \nu}={\rm diag}\{-\rho c^2,p,p,\cdots,p\}$ \cite{Jacobs1968.ApJ.153.661} and energy conservation is $\dot{\rho}+(\rho+p/c^2)\sum_{i=1}^{D-1}H_i=0$, where $H_i\equiv\dot{a}_i/a_i$. Our convention of the action is consistent with \cite{Fernandes2020.PRD.102.024025} and the gravitational field equation is \cite{Fernandes2020.PRD.102.024025}
\begin{equation}\label{eq:02}
  G_{\mu\nu}+\Lambda g_{\mu\nu}-\tilde{\alpha}H_{\mu\nu}=\kappa T_{\mu\nu},
\end{equation}
where $H_{\mu\nu}$ is given by Eq. (29) in \cite{Fernandes2020.PRD.102.024025}. The equation of motion for the scalar field is given by Eq. (30) in \cite{Fernandes2020.PRD.102.024025}. Substituting the 4-dimensional Bianchi type I metric into Eq. (\ref{eq:02}), we obtain
\begin{subequations}\label{eq:03}
\begin{align}
  &-\kappa\rho c^2=\Lambda-\frac{1}{c^2}(H_1H_2+H_1H_3+H_2H_3)\nonumber\\
  &\quad+\frac{\tilde{\alpha}}{c^4}\left[3\dot{\phi}^4+4\dot{\phi}^3(H_1+H_2+H_3)\right.\nonumber\\
  &\quad\left.+6\dot{\phi}^2(H_1H_2+H_1H_3+H_2H_3)+12\dot{\phi}H_1H_2H_3\right],\label{eq:03a}\\
  &\kappa p=\Lambda-\frac{1}{c^2}(A_i+A_j+H_iH_j)
  +\frac{\tilde{\alpha}}{c^4}\left[-\dot{\phi}^4+4\dot{\phi}^2\ddot{\phi}\right.\nonumber\\
  &\quad+2\dot{\phi}^2(A_i+A_j+H_iH_j)+4\dot{\phi}\ddot{\phi}(H_i+H_j)\nonumber\\
  &\quad\left.+4\dot{\phi}(H_iA_j+H_jA_i)+4\ddot{\phi}H_iH_j\right],\ {\rm for\ }ij=23,13,12,\label{eq:03b}
\end{align}
\end{subequations}
where $A_i\equiv\ddot{a}_i/a$. The equation of motions for the scalar field gives
\begin{subequations}\label{eq:04}
\begin{align}
  0&=3\dot{\phi}^2\ddot{\phi}+2\dot{\phi}\ddot{\phi}(H_1+H_2+H_3)\nonumber\\
  &+\ddot{\phi}(H_2H_3+H_1H_2+H_1H_3)+\dot{\phi}^3(H_1+H_2+H_3)\nonumber\\
  &+\dot{\phi}^2(A_1+A_2+A_3)+2\dot{\phi}^2(H_2H_3+H_1H_3+H_1H_2)\nonumber\\
  &+\dot{\phi}(H_2A_3+H_3A_2+H_1A_3+H_3A_1+H_1A_2+H_2A_1)\nonumber\\
  &+3\dot{\phi}H_1H_2H_3+H_1H_2A_3+H_1A_2H_3+A_1H_2H_3,\label{eq:04a}\\
  &=\frac{\dx[(\dot{\phi}+H_1)(\dot{\phi}+H_2)(\dot{\phi}+H_3)]}{\dx t}\nonumber\\
  &+(\dot{\phi}+H_1)(\dot{\phi}+H_2)(\dot{\phi}+H_3)(H_1+H_2+H_3).
\end{align}
\end{subequations}
Eq. (\ref{eq:04}) can be derived from Eq. (\ref{eq:03}) as we expected.

\section{The 4DEGB gravity}\label{sec:03}
In this section, we derive the cosmological evolution equations for the 4DEGB gravity. Hereafter we denote the set $S=\{1,2,\cdots,D-1\}$ and $S_i=\{j|j\in S,\ {\rm and}\ j\neq i\}$. Substituting the $D$-dimensional Bianchi type I metric into Eq. (\ref{eq:01}), we obtain
\begin{subequations}\label{eq:05}
\begin{align}
  &-\kappa\rho c^2=\Lambda-\frac{1}{c^2}\sum_{ij}H_iH_j-\frac{12\alpha}{c^4}\sum_{ijkl}H_iH_jH_kH_l,\\
  &\kappa p=\Lambda-\frac{1}{c^2}\sum_{j}A_j-\frac{1}{c^2}\sum_{jk}H_jH_k
  -\frac{4\alpha}{c^4}\sum_{jkl}H_jH_kA_l\nonumber\\
  &-\frac{12\alpha}{c^4}\sum_{jklm}H_jH_kH_lH_m,
  \ {\rm for\ }i=1,2,\cdots,D-1,
\end{align}
\end{subequations}
where the summation index $ij$ is the 2-combination of $S$, $ijkl$ is the 4-combination of $S$, $j$ is the 1-combination of $S_i$, $jk$ is the 2-combination of $S_i$, $jkl$ is the subset of 3 distinct elements of $S_i$ considering the order of $l$ \footnote{For example, $jkl\in\{234,423,342,235,523,352,245,524,$ $452,345,534,453\}$ for $D=6$ and $i=1$. The total number of the elements in this set equals to $3\times C_{D-2}^3$, where $C_n^m\equiv n!/[m!(n-m)!]$ is the binomial coefficient.}, $jklm$ is the 4-combination of $S_i$. The limit $D\rightarrow4$ with $\alpha=\tilde{\alpha}/(D-4)$ is not directly applicable to Eq. (\ref{eq:05}) because $H_{i|i>3}$ and $A_i|_{i>3}$ are unknown. To perform the dimensional-regularization approach, one must assume a specific symmetry. Here we study two kinds of symmetry.

\subsection{Symmetry w.r.t one axis}
In this subsection, we study the case where the extra dimensions are all symmetric with respect to one of the three spatial axes. We assume $H_{i|i>3}=H_n$ (and also $A_{i|i>3}=A_n$), where $n=1,2,$ or $3$. Performing the dimensional-regularization approach with this symmetry, Eq. (\ref{eq:05}) gives
\begin{subequations}\label{eq:06}
\begin{align}
  &-\kappa\rho c^2=\Lambda-\frac{1}{c^2}(H_1H_2+H_2H_3+H_1H_3)\nonumber\\
  &\ -\frac{12\tilde{\alpha}}{c^4}\left[-\frac{1}{4}H_n^4+\frac{1}{3}H_n^3(H_1+H_2+H_3)\right.\nonumber\\
  &\ \left.-\frac{1}{2}H_n^2(H_1H_2+H_2H_3+H_1H_3)+H_nH_1H_2H_3\right],\\
  &\kappa p=\Lambda-\frac{1}{c^2}(A_i+A_j+H_iH_j)
  -\frac{4\tilde{\alpha}}{c^4}\left[H_iH_jA_n\right.\nonumber\\
  &\ +H_iH_nA_j+H_nH_jA_i-(H_i+H_j)H_nA_n\nonumber\\
  &\ \left.-\frac{1}{2}H_n^2(A_i+A_j)+H_n^2A_n\right]
  -\frac{12\tilde{\alpha}}{c^4}\left[-\frac{1}{2}H_iH_jH_n^2\right.\nonumber\\
  &\ \left.+\frac{1}{3}(H_i+H_j)H_n^3-\frac{1}{4}H_n^4\right],\ {\rm for\ }ij=23,13,12.
\end{align}
\end{subequations}
Energy conservation can be rewritten as $\dot{\rho}+(H_1+H_2+H_3)\rho+(H_1p_{ij=23}+H_2p_{ij=13}+H_3p_{ij=12})/c^2=0$, and can be derived from Eq. (\ref{eq:06}) directly. As we mentioned in Sec. \ref{sec:01}, the above evolution equations depend on the choice of $n$ in $H_{i|i>3}=H_n$, i.e., the symmetry between the extra dimensions and the three spatial dimensions. This means the extra dimensions in the 4DEGB gravity can lead to physical influences. Therefore, we conclude that the extra dimensions have physical meaning and the 4DEGB gravity is essentially a higher-dimensional theory.

One can analytically verify that Eq. (\ref{eq:06}) is equivalent to Eq. (\ref{eq:03}) if $\dot{\phi}=-H_n$. Meanwhile, $\dot{\phi}=-H_n$ satisfies Eq. (\ref{eq:04}). Therefore, there is a correspondence between the 4DEGB gravity and the Horndeski gravity for this one-axis symmetry. This conclusion is consistent with the result obtained in \cite{Lin2020.2006.07913}, which analyzed a cylindrically symmetric metric. Note that this cylindrical metric has a similar mathematical structure compared with Bianchi type I metric, and its embedded symmetry is similar to the $H_{i|i>3}=H_n$ we adopted.

\subsection{Symmetry w.r.t three axes}
In this subsection, we study another symmetry choice. We assume $D=1+3N$, where $N$ is an integer. The symmetry is $H_{i+3n}=H_i$ for $i=1,2,3$ and $n=1,2,...,N-1$, i.e., $H_1=H_4=H_7=\cdots$, $H_2=H_5=H_8=\cdots$ and $H_3=H_6=H_9=\cdots$. In this case, the extra dimensions are symmetric with respect to the three spatial dimensions. The limit $D\rightarrow4$ corresponds to $N\rightarrow1$. Here we give the details of the derivation. With the assumed symmetry, Eq. (\ref{eq:05}) can be rewritten as
\begin{subequations}
\begin{align}
  -&\kappa\rho c^2
  =\Lambda-\frac{1}{c^2}(H_1H_2+H_2H_3+H_1H_3)\nonumber\\
  &-\frac{12\alpha}{c^4}\left[C_N^4(H_1^4+H_2^4+H_3^4)\right.\nonumber\\
  &+C_N^3C_N^1(H_1^3H_2+H_1^3H_3+H_2^3H_1+H_2^3H_3+H_3^3H_1\nonumber\\
  &+H_3^3H_2)+C_N^2C_N^2(H_1^2H_2^2+H_1^2H_3^2+H_2^2H_3^2)\nonumber\\
  &\left.+C_N^2C_N^1C_N^1(H_1^2H_2H_3+H_1H_2^2H_3+H_1H_2H_3^2)\right],\\
  \kappa&p
  =\Lambda-\frac{1}{c^2}(A_j+A_k+H_jH_k)\nonumber\\
  &-\frac{4\alpha}{c^4}\left[3C_{N-1}^3H_i^2A_i+3C_N^3(H_j^2A_j+H_k^2A_k)\right.\nonumber\\
  &+C_{N-1}^2C_N^1(2H_iH_jA_i+H_i^2A_j+2H_iH_kA_i+H_i^2A_k)\nonumber\\
  &+C_N^2C_{N-1}^1(2H_jH_iA_j+H_j^2A_i+2H_kH_iA_k+H_k^2A_i)\nonumber\\
  &+C_N^2C_N^1(2H_jH_kA_j+H_j^2A_k+2H_kH_jA_k+H_k^2A_j)\nonumber\\
  &\left.+C_{N-1}^1C_N^1C_N^1(H_iH_jA_k+H_iH_kA_j+H_jH_kA_i)\right]\nonumber\\
  &-\frac{12\alpha}{c^4}\left[C_{N-1}^4H_i^4+C_N^4(H_j^4+H_k^4)\right.\nonumber\\
  &+C_{N-1}^3C_N^1(H_i^3H_j+H_i^3H_k)\nonumber\\
  &+C_N^3C_{N-1}^1(H_j^3H_i+H_k^3H_i)+C_N^3C_N^1(H_j^3H_k+H_k^3H_j)\nonumber\\
  &+C_{N-1}^2C_N^2(H_i^2H_j^2+H_i^2H_k^2)+C_N^2C_N^2H_j^2H_k^2\nonumber\\
  &+C_{N-1}^2C_N^1C_N^1H_i^2H_jH_k\nonumber\\
  &\left.+C_{N-1}^1C_N^1C_N^2(H_iH_j^2H_k+H_iH_jH_k^2)\right], \nonumber\\
  &\qquad\qquad\qquad\quad {\rm for\ }ijk=123,213,312,
\end{align}
\end{subequations}
where we only keep the contribution of Einstein tensor in four-dimensional spacetime. The above partition is reasonable as one can easily verify
\begin{subequations}
\begin{align}
  &C_{3N}^4=3C_N^4+6C_N^3C_N^1+3C_N^2C_N^2+3C_N^2C_N^1C_N^1,\\
  &C_{3N-1}^3=C_{N-1}^3+2C_N^3+2C_{N-1}^2C_N^1+2C_N^2C_{N-1}^1\nonumber\\
  &\qquad\qquad+2C_N^2C_N^1+C_{N-1}^1C_N^1C_N^1,\\
  &C_{3N-1}^4=C_{N-1}^4+2C_N^4+2C_{N-1}^3C_N^1+2C_N^3C_{N-1}^1\nonumber\\
  &\qquad\qquad+2C_N^3C_N^1+2C_{N-1}^2C_N^2+C_N^2C_N^2\nonumber\\
  &\qquad\qquad+C_{N-1}^2C_N^1C_N^1+2C_{N-1}^1C_N^1C_N^2.
\end{align}
\end{subequations}
Then, taking $N\rightarrow1$ with $\alpha=\tilde{\alpha}/(D-4)$, we obtain
\begin{subequations}\label{eq:09}
\begin{align}
  -&\kappa\rho c^2
  =\Lambda-\frac{1}{c^2}(H_1H_2+H_1H_3+H_2H_3)\nonumber\\
  &\quad-\frac{\tilde{\alpha}}{c^4}\left[\frac{1}{3}(H_1^4+H_2^4+H_3^4)-\frac{2}{3}(H_1^3H_2+H_1^3H_3\right.\nonumber\\
  &\quad+H_2^3H_1+H_2^3H_3+H_3^3H_1+H_3^3H_2)\nonumber\\
  &\quad\left.+2(H_1^2H_2H_3+H_1H_2^2H_3+H_1H_2H_3^2)\right],\\
  \kappa&p=\Lambda-\frac{1}{c^2}(A_j+A_k+H_jH_k)
  -\frac{\tilde{\alpha}}{3c^4}\left[4H_i^2A_i\right.\nonumber\\
  &\quad-2(H_j^2A_j+H_k^2A_k)-4H_iA_i(H_j+H_k)\nonumber\\
  &\quad-2H_i^2(A_j+A_k)+4H_jH_k(A_j+A_k)\nonumber\\
  &\quad+2(H_j^2A_k+H_k^2A_j)\nonumber\\
  &\quad\left.+4(H_iH_jA_k+H_iH_kA_j+H_jH_kA_i)\right]\nonumber\\
  &\quad-\frac{\tilde{\alpha}}{c^4}\left[-H_i^4+\frac{1}{3}(H_j^4+H_k^4)+\frac{4}{3}H_i^3(H_j+H_k)\right.\nonumber\\
  &\quad\left.-\frac{2}{3}(H_j^3H_k+H_k^3H_j)-2H_i^2H_jH_k\right],\nonumber\\
  &\qquad\qquad\qquad\quad {\rm for\ }ijk=123,213,312.
\end{align}
\end{subequations}
Energy conservation can be derived from the above equations directly.

Is Eq. (\ref{eq:09}) equivalent to Eq. (\ref{eq:03}) for some specific scalar field configurations? Note that we also require this configuration to satisfy Eq. (\ref{eq:04}). It seems difficult to answer this question analytically. Here we try a numerical approach in which the scalar field configuration means the values of $\dot{\phi}$ and $\ddot{\phi}$. Without loss of generality, we assume $c=1$ and $\tilde{\alpha}=1$. We consider a special case where $\rho=0$ and $p=0$. For $H_1=1$, $H_2=1.3$ and $H_3=1.4$, Eq. (\ref{eq:09}) gives
\begin{subequations}
\begin{align}
  \Lambda&=11.0312,\\
  A_1&=1.6553,\\
  A_2&=1.4453,\\
  A_3&=1.2137.
\end{align}
\end{subequations}
Substituting the above results into Eq. (\ref{eq:03a}), we obtain two roots $\dot{\phi}=-1.1575$ and $\dot{\phi}=-0.9067$. For $\dot{\phi}=-1.1575$, Eq. (\ref{eq:04a}) gives $\ddot{\phi}=1.0968$. For $\dot{\phi}=-0.9067$, Eq. (\ref{eq:04a}) gives $\ddot{\phi}=-0.5618$. However, none of these solutions satisfy Eq. (\ref{eq:03b}). Therefore, we conclude that there is no scalar field configuration that satisfies Eq. (\ref{eq:04}) and simultaneously makes Eq. (\ref{eq:03}) and Eq. (\ref{eq:09}) equivalent. Back to the theory, there is no correspondence between the 4DEGB gravity and the Horndeski gravity for this three-axes symmetry.

\section{Conclusions}
In this paper, we highlight two issues about the 4DEGB gravity proposed in \cite{Glavan2020.PRL.124.081301}. One is the 4DEGB gravity is essentially a higher-dimensional theory. The other is, for the general case, the 4DEGB gravity is not equivalent to the Horndeski gravity proposed in \cite{Fernandes2020.PRD.102.024025}.

\textit{Note} --- See e.g. \cite{Arrechea2020.2004.12998,Gurses2020.2004.03390,Mahapatra2020.2004.09214,Shu2020.2004.09339} for other critical comments about the 4DEGB gravity.

Requests for the code (\texttt{Maple}) used to perform calculations in this paper should be addressed to SXT.

\section*{Acknowledgements}
SXT is supported by the Initiative Postdocs Supporting Program under Grant No. BX20200065. ZHZ is supported by the National Natural Science Foundation of China under Grant Nos. 11633001 and 11920101003, and the Strategic Priority Research Program of the Chinese Academy of Sciences under Grant No. XDB23000000.

%

\end{document}